\documentclass{article}
\usepackage{setspace}
\usepackage{lineno}
\usepackage{arxiv}
\usepackage{algorithm}
\usepackage{algpseudocode}
\usepackage[utf8]{inputenc} 
\usepackage[T1]{fontenc}    
\usepackage{hyperref}       
\usepackage{url}            
\usepackage{booktabs}       
\usepackage{amsfonts}       
\usepackage{nicefrac}       
\usepackage{microtype}      
\usepackage{lipsum}
\usepackage{graphicx}
\usepackage{amsmath}
\usepackage{rotating}
\usepackage{caption}
\usepackage{subcaption}
\usepackage[dvipsnames]{xcolor}
\definecolor{darkblue}{RGB}{0,0,139}
\definecolor{midnightblue}{RGB}{25,25,112}
\usepackage{caption}
\usepackage{natbib}
\usepackage{bm}
\setcitestyle{authoryear,open={(},close={)}} 
\usepackage[font={color=darkblue},figurename=Figure]{caption}
\graphicspath{ {./figures/} }
\usepackage{authblk}

\title{The impact of wettability on the co-moving velocity of two-fluid flow in porous media}
\author[a]{Fatimah Alzubaidi }
\author[b]{James E. McClure}
\author[c]{Håkon Pedersen}
\author[c]{Alex Hansen}
\author[d]{Carl Fredrik Berg}
\author[a]{Peyman Mostaghimi}
\author[a]{Ryan T. Armstrong}

\affil[a]{School of Minerals and Energy Resources Engineering, University of New South Wales, Sydney, Australia}
\affil[b]{National Security Institute, Virginia Tech, Blacksburg, Virginia, USA}
\affil[c]{PoreLab, Department of Physics, Norwegian University of Science and Technology, Trondheim, Norway}
\affil[d]{PoreLab, Department of Petroleum and Geoscience, Norwegian University of Science and Technology, Trondheim, Norway}

\begin{document}
\setcounter{tocdepth}{4}
\setcounter{secnumdepth}{4}
\maketitle

\begin{abstract}
The impact of wettability on the co-moving velocity of two-fluid flow in porous media is analyzed herein. The co-moving velocity, developed by Roy et al. (2022), is a novel representation of the flow behavior of two fluids through porous media. Our study aims to better understand the behavior of the co-moving velocity by analyzing simulation data under various wetting conditions. The simulations were conducted using the Lattice-Boltzmann color-fluid model and evaluated the relative permeability for different wetting conditions on the same rock. The analysis of the simulation data followed the methodology proposed by Roy et al. (2022) to reconstruct a constitutive equation for the co-moving velocity. Surprisingly, it was found that the coefficients of the constitutive equation were nearly the same for all wetting conditions. Based on these results, a simple approach was proposed to reconstruct the oil phase relative permeability using only the co-moving velocity relationship and water phase relative permeability. This proposed method provides new insights into the dependency of relative permeability curves, which has implications for the history matching of production data and solving the associated inverse problem. The research findings contribute to a better understanding of the impact of wettability on fluid flow in porous media and provide a practical approach for estimating relative permeability based on the co-moving velocity relationship, which has never been shown before.

\end{abstract}
\pagebreak

\section{Introduction}
The phenomenological two-phase extension of Darcy's law was proposed by Wyckoff and Botset 1936 \cite{wyckoff1936flow}. The conceptual idea was that during flow the presence of one fluid occludes the movement of the other fluid and vice versa. The concept of relative permeability captures this behaviour by defining 'how permeable' a medium is to a given fluid based on its saturation. The concept of 'how permeable' is measured relative to the permeability of the medium for single-phase flow. The prevalence and longevity of the two-phase extension of Darcy's law in the porous media community speaks to its usefulness and simplicity. Herein, we explore a thermodynamic framework provided by \cite{hansen2018relations} that is compatible with the two-phase extension of Darcy's law yet provides (1) new insights into the transport behaviour of immobile phases in porous media and (2) a new capability for constructing relative permeability curves that has never been shown before.  

The theory presented by \cite{hansen2018relations} provides a fresh perspective on the ``kinematics" of multiphase flow in the sense that the theory is only concerned with relationships between the velocities of the fluids. The fluid velocities are the seepage velocity, i.e. the average pore velocity,
of each fluid, $v_w$ and $v_n$, where the subscripts $w$ and $n$ refer to the more wetting and the less wetting fluid, respectively. We define the total seepage velocity as
\begin{equation}
    v_p=S_wv_w+S_nv_n,
    \label{AH-1}
\end{equation}
where $S_w+S_n=1$ are the saturations. The {\it co-moving velocity\/} is defined as 
\begin{equation}
    v_m=S_w\frac{dv_w}{dS_w}+S_n\frac{dv_n}{dS_w},
    \label{AH-2}
\end{equation}
see \cite{hansen2018relations,feder2022physics,roy2022co,pedersen2023parameterizations} for further details.
These two equations may be inverted to give
\begin{equation}
    \label{AH-3}
    v_w=v_p+S_n\left[\frac{dv_p}{dS_w}-v_m\right],
\end{equation}
and
\begin{equation}
    \label{AH-4}
     v_n=v_p-S_w\left[\frac{dv_p}{dS_w}-v_m\right].
\end{equation}
Hence, knowing the total seepage velocity $v_p$ and the co-moving velocity $v_m$, we find the seepage velocity for each fluid. The four equations constitute a two-way mapping between the pairs $(v_w,v_n)$ and $(v_p,v_m)$. 

By taking the derivative of the total seepage velocity $v_p$ with respect to the wetting saturation $S_w$ in Equation \ref{AH-1}, and combining the result with the definition of the co-moving velocity in Equation \ref{AH-2}, we find the expression
\begin{equation}
    \frac{dv_p}{dS_w} = v_w - v_n + v_m,
    \label{co}
\end{equation}
demonstrating that the co-moving velocity is related to the velocity difference between the two fluids. Intuitively, the flow of each fluid carries along with it some of the other fluid. In addition, the cluster structure of two immiscible fluids can interact with the porous medium. The degree to which these two effects occur is captured by the co-moving velocity. 

The co-moving velocity has been studied for various porous systems \cite{roy2022co} and was found to follow a simple functional form,
\begin{equation}
    v_m = av_0 + b \frac{dv_p}{dS_w}.
    \label{roy}
\end{equation}
   
From Equations \ref{co} and \ref{roy},
\begin{equation}
    v_n - v_w = av_0 + (b-1)\frac{dv_p}{dS_w},
    \label{diff}
\end{equation}
the linear relationship is seen as a consequence of the velocity difference between the flowing fluids. The co-moving velocity relates the seepage velocities of the two competing fluids within a given porous medium.  

The linear relationship (Equations \ref{diff} or \ref{roy}) has been demonstrated for a wide range of Capillary numbers and Mobility ratios \cite{roy2022co}. Remarkably, the co-moving velocity relationship remains linear even when the total seepage velocity relationship becomes non-linear \cite{roy2022co}. \cite{hansen2023statistical} showed why $dv_p/dS_w$ is the natural variable for $v_m$.  However, 
it remains an open question as to why the co-moving velocity takes such a simple affine form. The co-moving velocity was first formulated in a thermodynamic-like description of two-phase flow in porous media \cite{hansen2018relations}. In the mathematical structure of thermodynamics, the linear relation of the co-moving velocity is possible, but physically such a simple relationship is not required and not necessarily expected. We speculate that the relationship is a consequence of partitioning the system into two separate fluid systems. However, further theoretical work is required to bring any validity to this speculation. Nevertheless, exploring the co-moving velocity relationship with experimental and simulation data will provide further insights that can guide the development of the theory, which is the aim of this paper. 

The wetting state of a porous media is a major factor that influences the geometric state of the fluids and thus, relative permeability \cite{anderson1986wettability,anderson1987wettability}. Studying the co-moving velocity under different wetting conditions is a logical step to unravelling the linearity of the co-moving velocity relationship. At the pore scale, the geometrical arrangement of fluids depend on the local wetting conditions, which impacts the surface area coverage \cite{garfi2020fluid}, Gaussian curvature \cite{lin2019minimal}, and mean curvature \cite{lin2018imaging} of the fluid clusters. The energy necessary to move an interface also becomes less when the surface energies between the fluids and solid are similar, i.e., near a contact angle of $90^{\deg}$ \cite{armstrong2021multiscale}. Therefore the propensity for interfacial rearrangements and intermittency increases for intermediate-wet conditions resulting in complex interactions between the flowing fluids \cite{reynolds2017dynamic,zou2018experimental,scanziani2020dynamics}. It is therefore reasonable to assume that wettability would impact the co-moving velocity relationship in unexpected ways. 

Herein, we investigate the impact of wettability on the constitutive relationship for the co-moving velocity, Equation \ref{roy}. Firstly, multiphase flow simulations for a range of different wetting conditions are analysed to investigate the co-moving velocity relationship. Simulation data are based on two separate rock types under a range of different wetting conditions, which have been published elsewhere \cite{PhysRevFluids.8.064004}. Secondly, we substantiate our result based on simulation data by using experimental core flooding data from \cite{zou2018experimental}. Lastly, we demonstrate a practical implication by reconstructing full relative permeability curves when only the relative permeability of a single phase and its co-moving velocity relationship are known.

\section{Materials and Methods}
Firstly, we will explain the theoretical developments used in this paper. The simulation relative permeability data will then be explained followed by the experimental data used to substantiate our results. 

\subsection{Co-moving Velocity and Relative Permeability}
Two-phase Darcy law for the seepage (i.e., average pore) velocities when there are no saturation gradients present, is expressed as, 
\begin{equation}
    v_i = \frac{k_{ri} \mathsf{K} }{\phi S_i\mu_i} |\nabla p|,
\label{darcy}
\end{equation}

where the usual minus sign is combined with the negative gradient to produce a positive number, expressed as $|\nabla p|$. Furthermore, $k_{ri}$ is the relative permeability of phase $i$, $\mu_i$ is the viscosity of phase $i$, $\phi$ is the porosity, $\mathsf{K}$ is absolute permeability, and $p$ is pressure.

The co-moving velocity can be directly determined from relative permeability. From Equations \ref{co} and \ref{darcy},
\begin{equation}
v_{m}/v_{0} = \frac{d(v_{p}/v_{0})}{dS_{w}}+\mu_{w} \left[ \frac{k_{rn}}{\mu_{n} S_{n}}-\frac{k_{rw}}{\mu_{w} S_{w}}\right].
\label{eq:vm}
\end{equation}

Where 
\begin{equation}
v_{p}/v_{0} = \mu_{w}\left[ \frac{k_{rw}}{\mu_{w}}+\frac{k_{rn}}{\mu_{n}}\right].
\label{eq:vp}
\end{equation}

We define a velocity scale that must be positive,

\begin{equation}
v_{0} = \frac{\mathsf{K}}{\mu_{w}\phi} |\nabla p|.
\label{eq:vo}
\end{equation} 

Our analysis relies on the produced relationship between $v_{p}/v_{0}$ versus $S_{w}$, $v_{m}/v_{0}$ versus $S_{w}$, and $v_{m}/v_{0}$ versus $d(v_{p}/v_{0})/dS_{w}$. The latter represents the constitutive relationship for the co-moving velocity as given in Equation \ref{roy}.

Lastly, the Corey model for relative permeability will be used in Section \ref{prediction} of the results. The Corey model is commonly used to approximate the functional relationship between the relative permeability and saturation,
\begin{eqnarray}
    k_{rn} &=& k_{rn,max}\Bigg( \frac{1-S_w-S_{nr}}{1-S_{iw}-S_{nr}}\Bigg) ^{\eta_o} \\
    k_{rw} &=& k_{rw,max} \Bigg( \frac{S_w-S_{iw}}{1-S_{iw}-S_{nr}}\Bigg) ^{\eta_w}\;,
    \end{eqnarray}
where $S_{iw}$ is irreducible wetting phase saturation, $S_{nr}$ is residual non-wetting phase saturation, $k_{rn,max}$ is the maximum non-wetting phase relative permeability, and $k_{rw,max}$ is the maximum wetting phase relative permeability. 

\subsection{Pore-scale Simulations}
Fluid flow simulations were conducted using a parallel implementation of the colour-gradient Lattice Boltzmann method. The code has been documented and validated elsewhere and allows for spatially varied wetting conditions on a per-voxel basis \cite{mcclure2021lbpm,armstrong2016beyond}. 

Two different sandstone samples were evaluated. The North Sea sandstone had porosity and permeability of 23\% and 640 mD, respectively. The Bentheimer sandstone had porosity and permeability of 24\% and 1.3 D, respectively. Both samples were imaged with X-ray microcomputed tomography (micro-CT) to generate a simulation domain. 

The North sea sandstone simulation domain was generated by coupling micro-CT images with scanning Electron Microscopy Energy Dispersive Spectroscopy (SEM-EDS) followed by Quantitative Evaluation of Materials by Scanning Electron Microscopy (QEMSCAN) software. The result provided a full three-dimensional image of the pore-structure and mineralogy at 2.3 $\mu$m micrometer resolution. The total image size was $1000\times1000\times750$ voxels. For simplicity the minerals identified by QEMSCAN were categorized into three main groups: quartz, carbonate, and clay for wettability assignment. Full details are published in \cite{PhysRevFluids.8.064004}. 

The Bentheimer sandstone simulation domain was generated by micro-CT only, and thus did not contain any mineralogical information. The pore structure was imaged at 1.66 $\mu$m micrometer resolution. The resulting image size was $900\times900\times1600$ voxels. These simulations results are unique to this publication but utilize the same method as published in \cite{mcclure2021lbpm, PhysRevFluids.8.064004}.

Wettability was controlled by changing the contact angle, which at equilibrium relates to various surface energies in the two-fluid system as defined by Young’s equation. These surface energies can be assigned to the solid surface domain on a per voxel basis \cite{mcclure2021lbpm}. The wettability was distributed four ways.
\begin{enumerate}
\item Homogeneous Wet: same condition over entire grain surface.
\item Corner Wet: oil is injected based on a morphological approach to achieve irreducible water saturation. Parts of the grain surface that touch oil get their own wetting condition while corners remain water wet.
\item Heterogeneous Wet: each mineral type gets a different wettability. 
\item Heterogeneous-Corner Wet: combination of 2 and 3.
\end{enumerate}
For the North Sea sandstone all wettability types are used. For the Bentheimer sandstone only Type 2 wettability was used.

The domain wettability was defined by summation of the cosine of the contact angles for each mineral/fluid/fluid pair, determined as 
\begin{equation}
W = 
(\gamma_{qn}-\gamma_{qw})/\gamma_{wn} \phi_{q} +
(\gamma_{kn}-\gamma_{kw})/\gamma_{wn} \phi_{k} +
(\gamma_{cs}-\gamma_{cs})/\gamma_{wn} \phi_{c},
\label{eq:wetindex}
\end{equation}
where $\phi_{q}$, $\phi_{k}$ and $\phi_{c}$ are the solid voxel fractions of quartz, clay, and carbonate, respectively. This metric provides a scale from strongly water-wet ($W = 1.0$) to strongly oil-wet ($W = -1.0$).

The simulated fluids had equal viscosity and Capillary number was approximately $10^-5$ for all simulations. The simulations were conducted in steady state mode whereby any phase leaving the outlet was introduced to the inlet to keep fractional flow and saturation constant until the simulation converged \cite{mcclure2021lbpm}. Such requirements do not allow for the development of a saturation gradient nor saturation fluctuations. Convergence criteria was based on Capillary number stabilisation \cite{mcclure2021lbpm}.

\subsection{Experimental Data}
The experimental data was based on the work of Zou et al. \cite{zou2018experimental} the full details are reported therein. Only the salient points are provided below. 

Core flooding studies were conducted using a mini-core plug of Bentheimer sandstone. The sample had a porosity and permeability of 24\% and 2.54 D, respectively. The core was 18.5 mm in length with a diameter of 10.2 mm. The fluids were decane and water. Relative permeability was first measured in the core under its clean water-wet state. The core was then treated with octadecytrichlorosilane (OTS) while frozen with connate water. OTS rendered the exposed core surfaces oil-wet with an estimated contact angle of $115^{\deg}$. Relative permeability was then measured on the mixed-wet core. 

Relative permeability was measured using the steady state method \cite{mcphee2015preparation}. Capillary end-effect was accounted for by history matching the steady state data. An independently measured capillary pressure versus saturation curve, porosity, and absolute permeability were used as known values for the history matching. Saturation profiles along the core were determined from micro-CT images of the core at each steady state condition.    

\section{Results and Discussion}
Firstly, we will present the co-moving velocity relationships for the simulated and experimental determined relative permeability data. We will then present a method to predict the oil phase relative permeability based on a simplified co-moving velocity relationship and the corresponding water phase relative permeability. 

\subsection{Co-moving Velocity Relationship}
\begin{figure}
     \centering
     \begin{subfigure}[b]{0.45\textwidth}
         \centering
         \includegraphics[width=\textwidth]{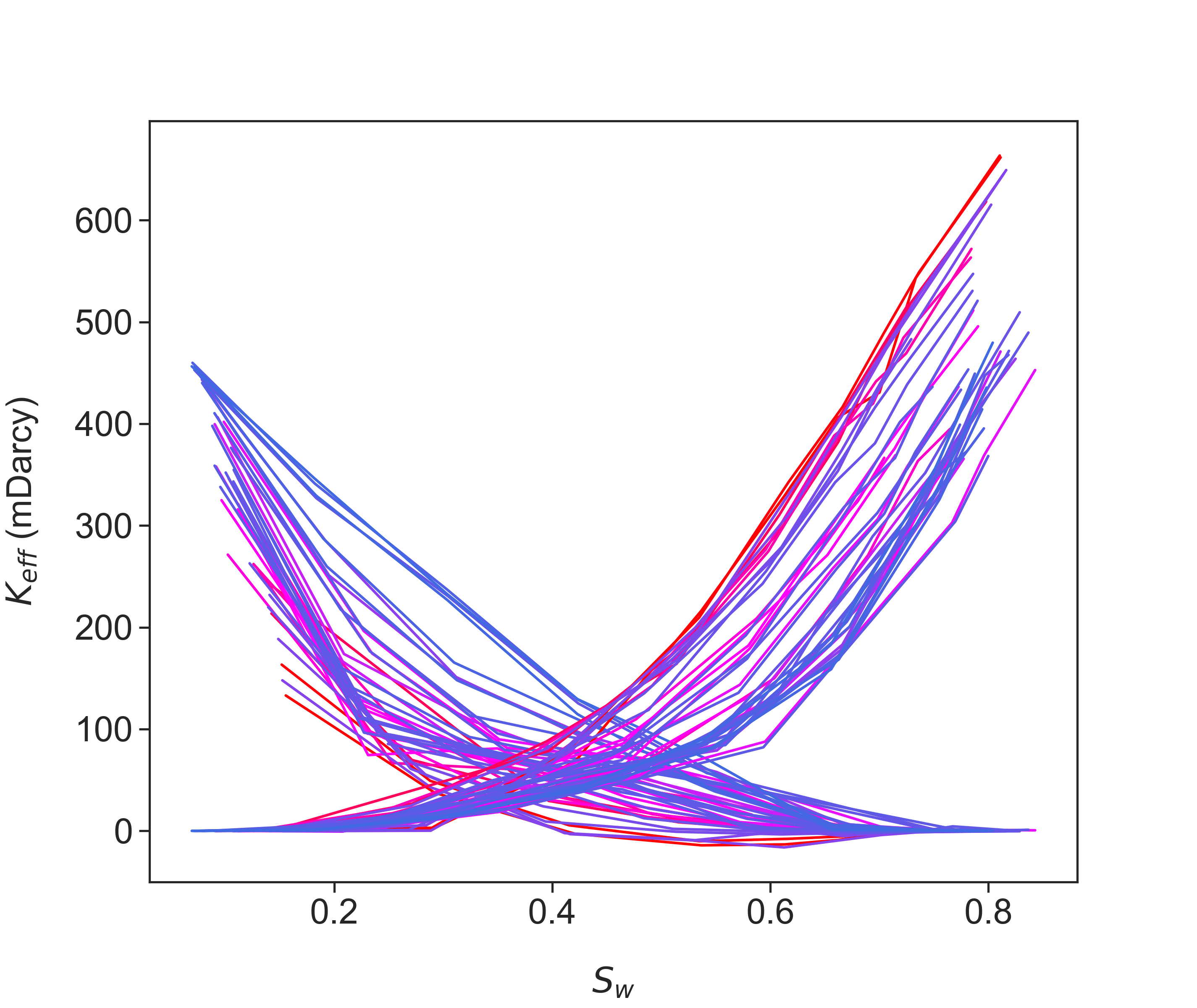}
         \caption{North Sea Sandstone}
         \label{fig:North}
     \end{subfigure}
     \hfill
     \begin{subfigure}[b]{0.53\textwidth}
         \centering
         \includegraphics[width=\textwidth]{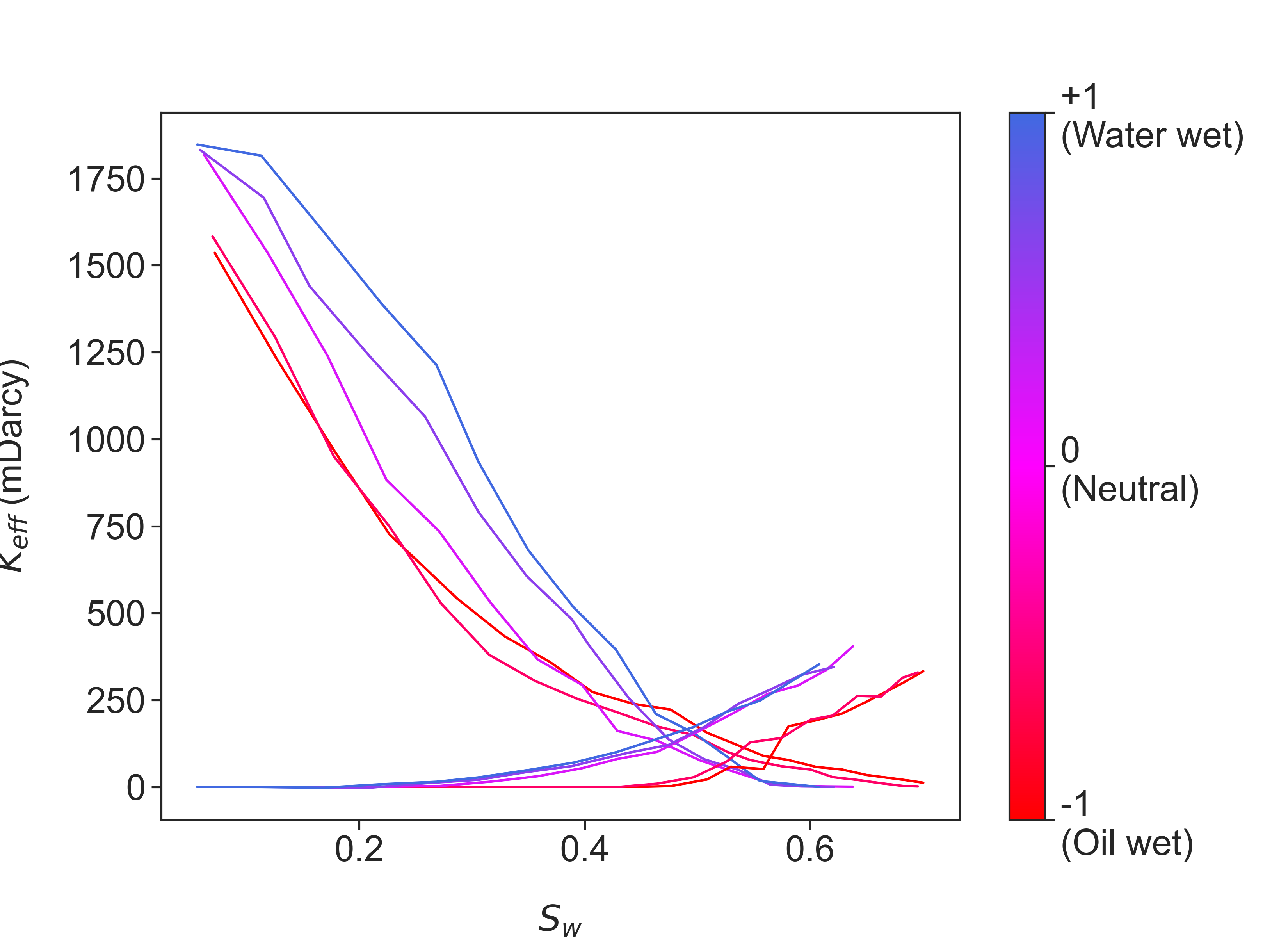}
         \caption{Bentheimer sandstone}
         \label{fig:bent}
     \end{subfigure}
     \hfill
      \hfill
     \begin{subfigure}[b]{0.49\textwidth}
         \centering
         \includegraphics[width=\textwidth]{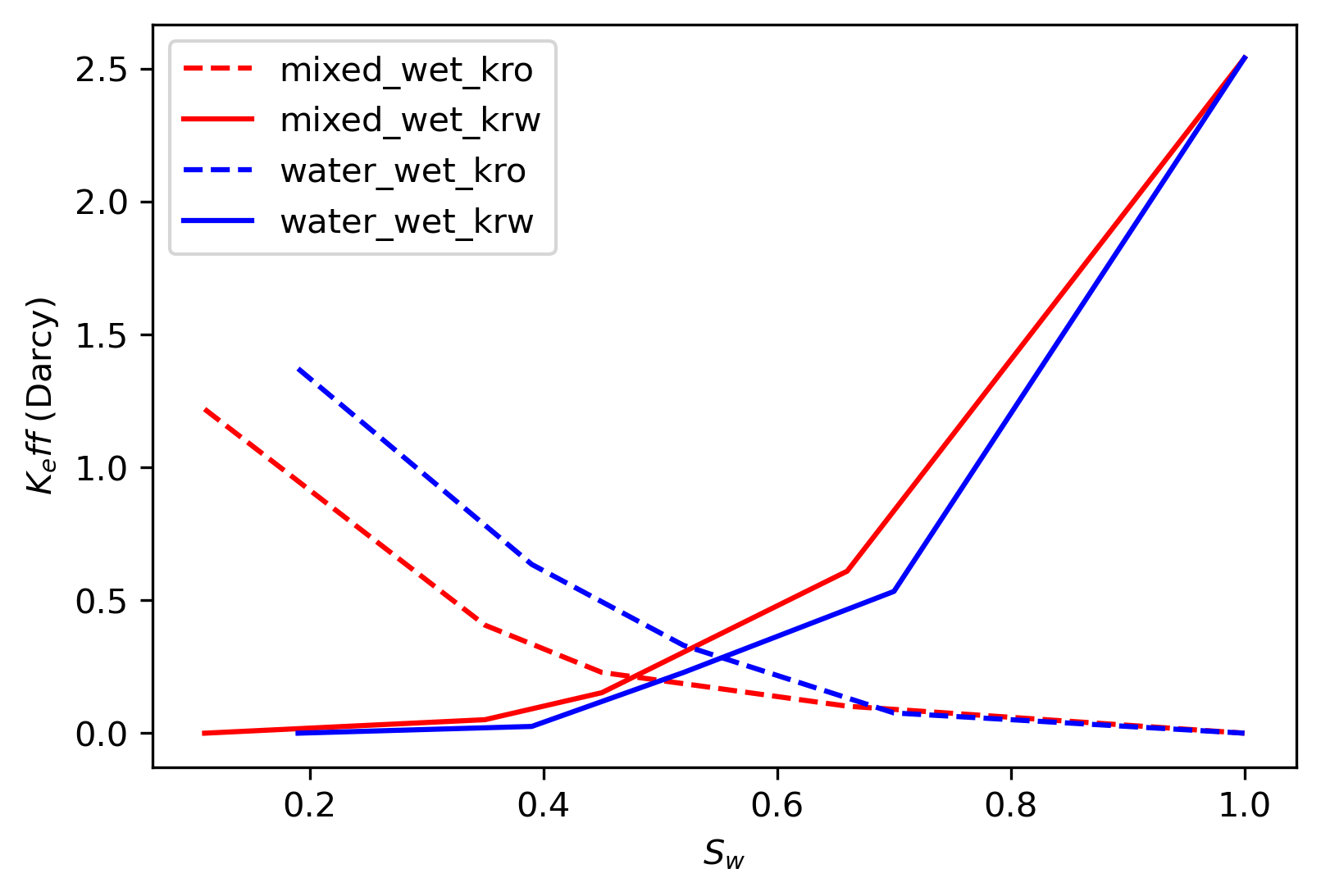}
         \caption{Bentheimer Experimental}
         \label{fig:bent}
     \end{subfigure}
     \hfill
        \caption{Relative permeability curves for a range of different wetting conditions. (a) and (b) are based on simulation data. (c) is experimentally measured steady state relative permeability.}
        \label{fig:rel_perm}
\end{figure}

All relative permeabilities are shown in Figure \ref{fig:rel_perm}. In total, 41 different wettability cases were simulated for the North Sea sandstone while five different wettability cases were simulated for the Bentheimer sandstone. In addition, two different wettability cases were measured experimentally for Bentheimer sandstone. For all data, the relative permeability curves provided the expected trend whereby oil phase relative permeability increases as the rock becomes more water wet while the water phase permeability decreases. Albeit the curves differ in terms of various other aspects the general trends are consistent. 

\begin{figure}[H]
\includegraphics[width=\textwidth]{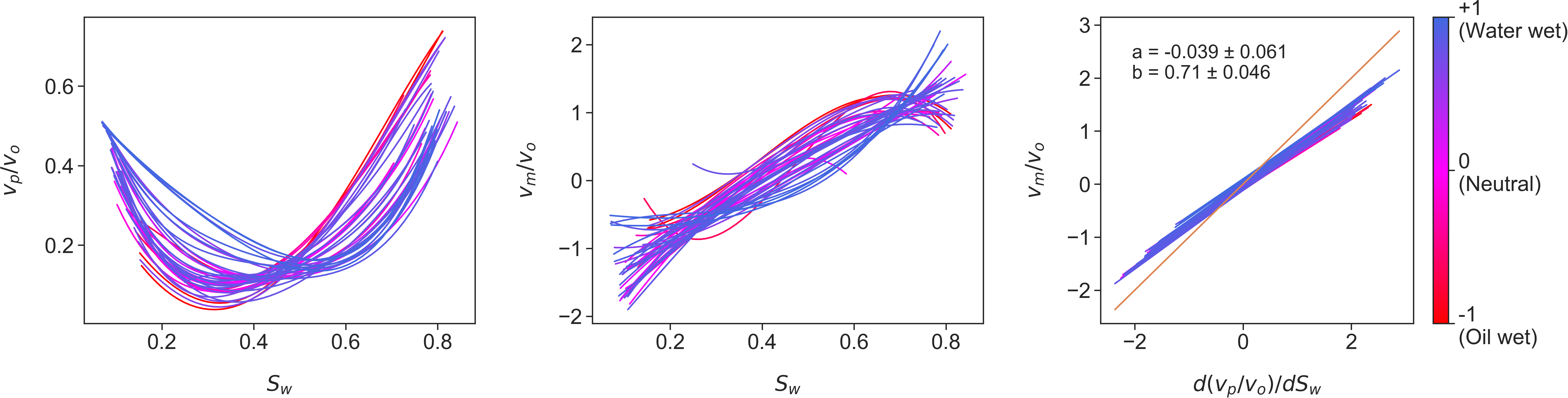}
\caption{Derived relationships for the North Sea sandstone based on 41 different wetting conditions.}
\label{results:all}
\end{figure}

\begin{figure}
     \centering
     \begin{subfigure}[b]{1.0\textwidth}
         \centering
         \includegraphics[width=\textwidth]{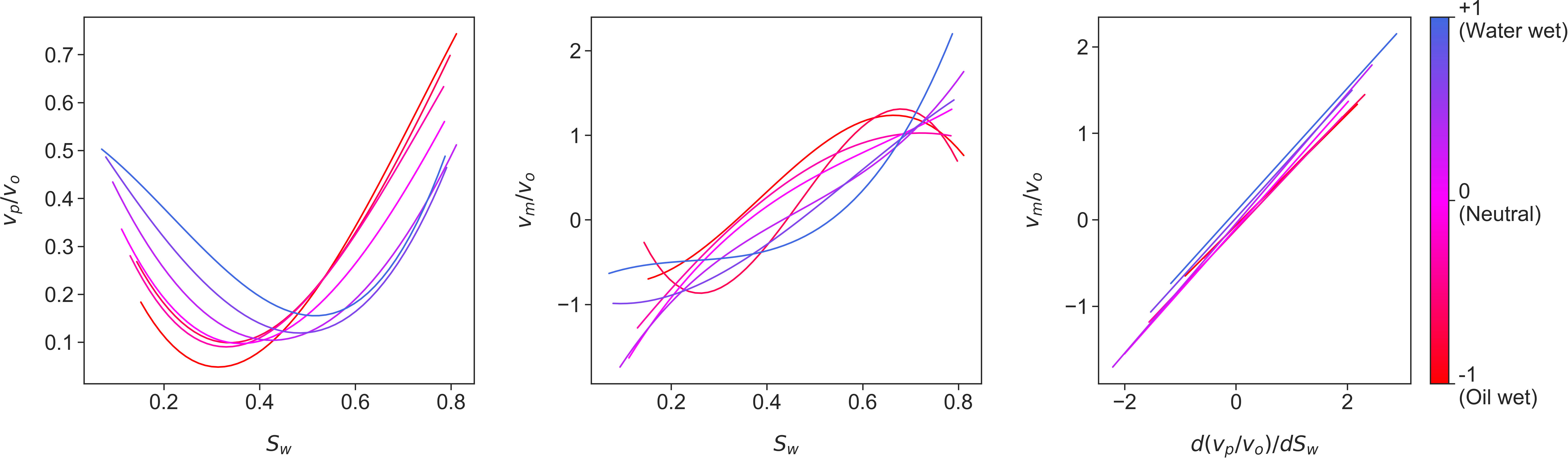}
         \caption{Homogeneous Wet}
         \label{fig:homo}
     \end{subfigure}
     \hfill
     \begin{subfigure}[b]{1.0\textwidth}
         \centering
         \includegraphics[width=\textwidth]{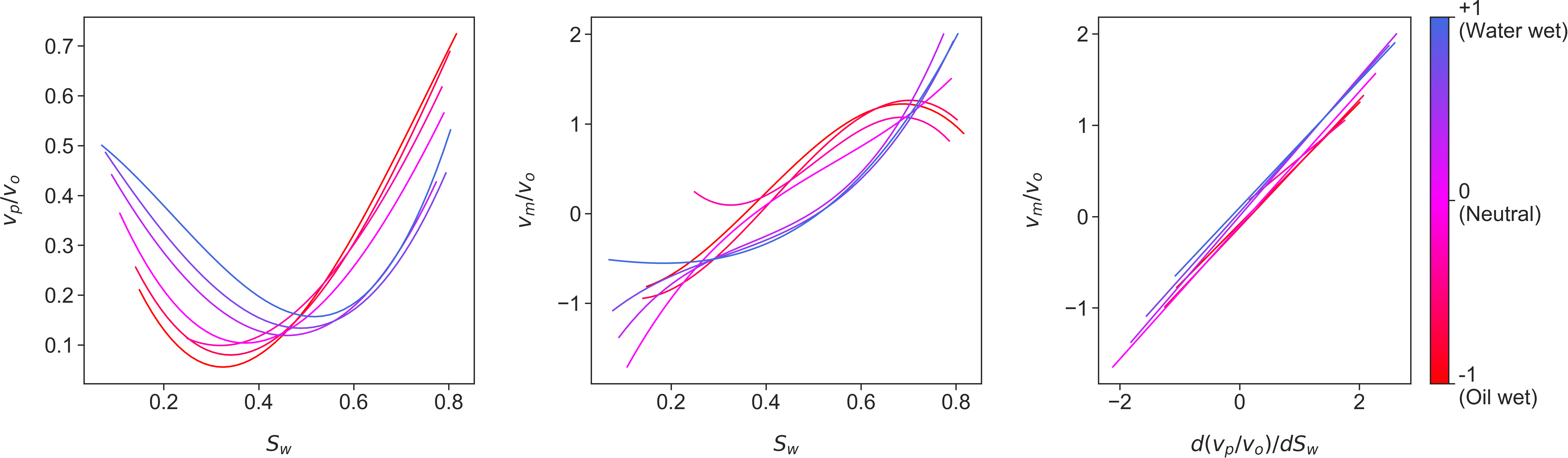}
         \caption{Corner Wet}
         \label{fig:mix}
     \end{subfigure}
     \hfill
     \begin{subfigure}[b]{1.0\textwidth}
         \centering
         \includegraphics[width=\textwidth]{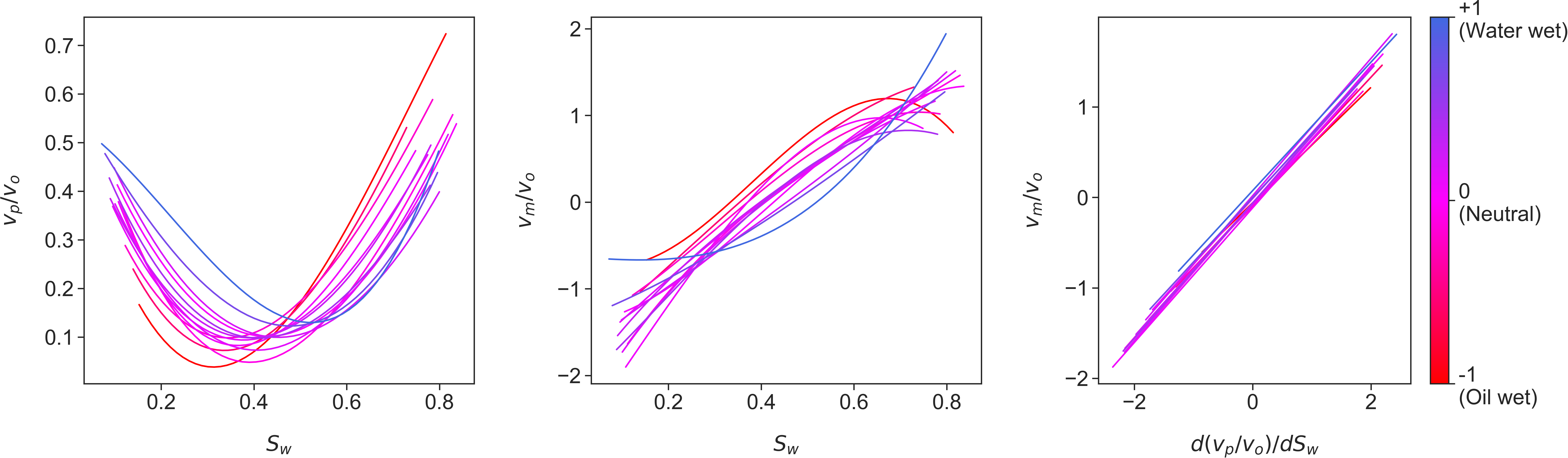}
         \caption{Heterogeneous Wet}
         \label{fig:hete}
     \end{subfigure}
     \hfill
     \begin{subfigure}[b]{1.0\textwidth}
         \centering
         \includegraphics[width=\textwidth]{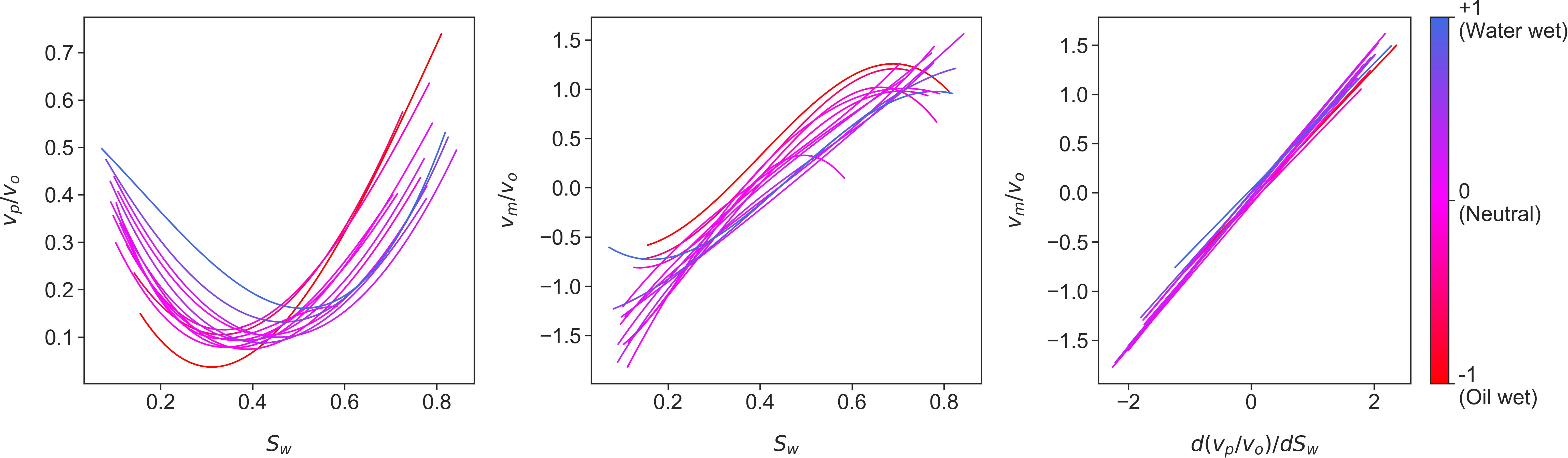}
         \caption{Heterogeneous-Corner Wet}
         \label{fig:het_mix}
     \end{subfigure}
        \caption{Derived correlations for North Sea Sandstone with results for each wetting condition.}
        \label{fig:Wetting}
\end{figure}

The derived correlations for the North Sea data are presented in Figure \ref{results:all}, see also Figure \ref{fig:Wetting}. The results indicate that $v_{p}/v_{0}$ versus $S_{w}$, and $v_{m}/v_{0}$ versus $S_{w}$ varied among the different wetting conditions. In general, the curves shifted towards the right (higher $S_{w}$) for more water-wet cases, which is due to their associated relative permeability curves that also exhibit a similar trend for water-wet conditions. Remarkably $v_{m}/v_{0}$ versus $d(v_{p}/v_{0})/dS_{w}$ provided a linear relationship that was nearly the same for all wettability conditions. The linear relationship was consistent for 40 of the 41 different wetting states tested.

\begin{figure}[H]
\includegraphics[width=\textwidth]{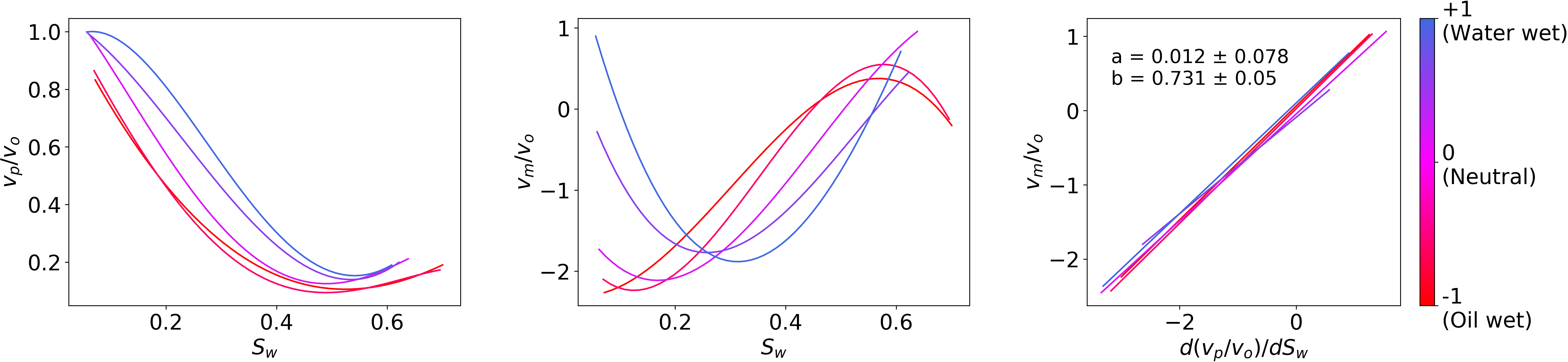}
\caption{Derived correlations for Bentheimer sandstone with results from 5 different wetting conditions.}
\label{results:all_bent}
\end{figure}

The derived correlations for the Bentheimer sandstone simulations are presented in Figure \ref{results:all_bent}. Overall, the $v_{p}/v_{0}$ versus $S_{w}$, and $v_{m}/v_{0}$ versus $S_{w}$ relationships varied to that observed for the North Sea sandstone. However, as observed with the North Sea sample, the linear relationship between $v_{m}/v_{0}$ and $d(v_{p}/v_{0})/dS_{w}$ was not influenced by the wetting condition. These results confirm with an independent sample that wettability has little to no effect on the linear relationship for the co-moving velocity. 

\begin{figure}[H]
\includegraphics[width=\textwidth]{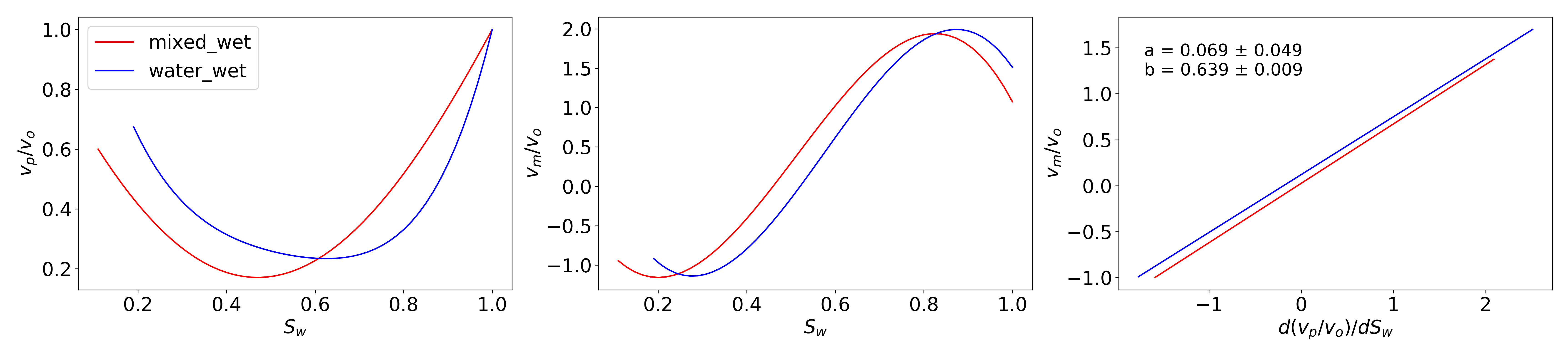}
\caption{Derived correlations based on the Bentheimer steady-state experimentally measured relative permeability data.}
\label{results:exp_bent}
\end{figure}

The derived correlations for the Bentheimer sandstone experimental measurements are presented in Figure \ref{results:exp_bent}. As observed with both simulation data sets, the wetting condition had little to no effect on the linear relationship between $v_{m}/v_{0}$ and $d(v_{p}/v_{0})/dS_{w}$. These results confirm that the observed consistent linear relationship is of physical origin and not caused by the numerical method implemented. 

 The average coefficients $a$ and $b$ for the simulated North Sea and Bentheimer sandstone samples are reported in Table \ref{coef}. Given the standard deviation for the coefficient, the coefficient values for both rock samples are considerably similar. The samples were similar in respect to permeability and porosity in addition to both samples being sandstone. It appears that the differences in rock structure for these samples was not strong enough to clearly observe any discernible differences in the co-moving velocity relationship.
 
\begin{table}[H]
\centering
\caption{Coefficients $A$ and $B$ for the co-moving velocity relationship. The reported coefficients are the average for all simulated wetting states and associated errors are based on the standard deviation.}
\begin{tabular}{lll}
\textbf{}    & $a$ & $b$ \\ \hline
North Sea Sandstone                & $-0.039 \pm 0.061 $          & $0.71 \pm 0.046$ \\
Bentheimer Sandstone       & $0.012 \pm 0.078  $       & $0.731 \pm 0.050$ \\ \hline
\end{tabular}
\label{coef}
\end{table}

\subsection{Relative Permeability Prediction} \label{prediction}
An essential result of this work is that wettability had limited influence on the co-moving velocity relationship. Therefore, once this relationship is known it can be applied to any given wetting case or rock provided that the pore structure (rock type) remains similar. 

To simplify the co-moving velocity relationship, we propose
\begin{equation}
    v_m = b \frac{d v_p}{d S_w},
    \label{new}
\end{equation}
given that $a$ was nearly zero for all tested cases. This provides a condition (or constraint) on how one relative permeability curve must change in relation to another. Therefore, given one relative permeability curve and $b$ for Equation \ref{new}, the other paired relative permeability can be predicted. This we see from Eqs. \ref{eq:vm} and \ref{eq:vp}: knowing one of the relative permeabilities, we can turn these two equations into a differential equation for the other one. 

In the following, we proceed along a path that does not require solving a differential equation. We start with a Corey relationship for the oil phase relative permeability with an unknown Corey exponent, $\eta_o$. This Corey relationship is paired with the simulated known water phase relative permeability. The co-moving velocity relationship can then be determined using this paired data (using the approach explained in Section 2) to provide $b_{predict}$. We then search $\eta_o$ over a reasonable range of Corey Exponents (1 to 10) to find $\min |b_{predict} - b_{NS}|$, where $b_{NS}$ is the average coefficient value ($b$) for the North Sea data (see Table \ref{coef}.

\begin{figure}[H]
\includegraphics[width=\textwidth]{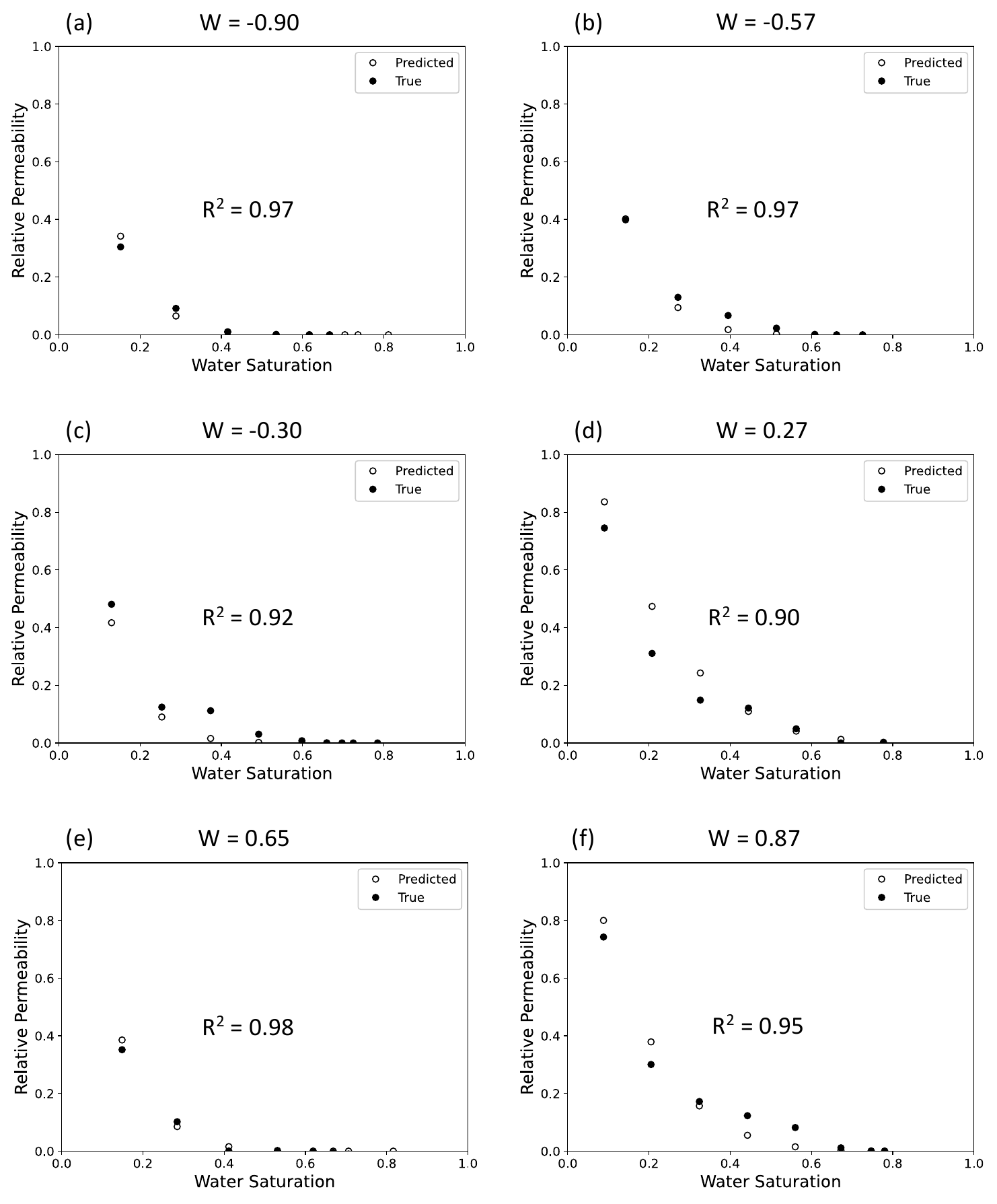}
\caption{Predicted versus true relative permeability curves for the oil phase. Predictions are based on the simulated water phase relative permeability and a single average co-moving velocity relationship for the North Sea sandstone.}
\label{results:predict}
\end{figure}

Six simulation cases from the North Sea Sandstone data that range the entire span of simulated wettabilities are presented in Figure \ref{results:predict}. For each case the predicted oil phase relative permeability curves are nearly equivalent to the actual simulation data. The wettability index, $W$, is reported for each tested case. The prediction error is likely associated with two factors. (1) The co-moving velocity relationship was simplified by considering a y-intercept of zero for the general model. (2) The maximum oil phase relative permeability, $k_{rn,max} = 1$, was assumed to be constant for all case. $k_{rn,max}$ could indeed be used as an additional unknown fitting parameter but the aim was to demonstrate the simplest approach possible.

\section{Conclusion}
Our main result is that the co-moving velocity relationship remains linear for a wide range of wetting conditions. In addition, the wetting state had little to no effect on the coefficients for the co-moving velocity relationship. Our results were consistent for both experimental and simulation data. Other works have also reported a simple affine form for the co-moving velocity based on pore network modelling. Therefore, a wide range of different capillary numbers, fluid properties, rock types, and wetting conditions all provide a linear relationship. Our work differs from previous works by two factors. (1) We demonstrate a universal behaviour for different wetting states. (2) We demonstrate a workflow to predict complete relative permeability curves when only one set of relative permeabilities are measured and the co-moving velocity relationship is known.

The universality of the linear relationship for the co-moving velocity is a remarkable result. Total seepage velocity and the co-moving velocity are a coupled pair that contain the same information as the coupled wetting and nonwetting seepage velocities, i.e., $(v_p,v_m) \leftrightarrow (v_w,v_n)$. Therefore, relative permeability within the traditional two-phase extension of Darcy's law can be determined by measuring only the total seepage velocity and average pressure gradient once the co-moving velocity relationship is known for a given rock type. Such measures are experimentally easier than measuring the seepage velocity of each immiscible fluid. Based on our results new laboratory workflows can be envisaged when determining the relative permeability of a rock core under different wetting conditions. History matching approaches used to determine relative permeability would also benefit from the additional constraint provided by the co-moving velocity relationship since it links the behaviour of the two relative permeability curves. Overall, our results show that the representation of two fluid flow based on the co-moving velocity is simpler and more universal than expected. This suggests that the alternative representation may have advantages with respect to the conventional representation, which warrants further research. 

\section{Acknowledgment}
\label{sec:Acknowledgement}
R. T. A. acknowledges Australian Research Council Future Fellowship (FT210100165) and Discovery (DP210102689). This work was furthermore supported by the Research Council of Norway through its Center of Excellence funding scheme, project number 262644 (HP, AH and CFB). This research used resources of the Oak Ridge Leadership Computing Facility, which is a DOE Office of Science User Facility supported under Contract DE-AC05-00OR22725.

\section{Data}
Upon publication, all data and code will be made available at https://github.com/fatimahgit.

\bibliography{reference} 
\end{document}